\documentclass[sigconf]{acmart}

\copyrightyear{2023}
\acmYear{2023}
\setcopyright{acmlicensed}
\acmConference[CSCW '23 Companion]{Computer Supported Cooperative Work and Social Computing}{October 14--18, 2023}{Minneapolis, MN, USA}
\acmBooktitle{Computer Supported Cooperative Work and Social Computing (CSCW '23 Companion), October 14--18, 2023, Minneapolis, MN, USA}
\acmPrice{15.00}
\acmDOI{10.1145/3584931.3606996}
\acmISBN{979-8-4007-0129-0/23/10}

\AtBeginDocument{%
  \providecommand\BibTeX{{%
    \normalfont B\kern-0.5em{\scshape i\kern-0.25em b}\kern-0.8em\TeX}}}

\begin{document}

\title{The Synthesis Lab: Empowering Collaborative Learning in Higher Education through Knowledge Synthesis}

\author{Xinran Zhu}
\email{xrzhu@upenn.edu}
\orcid{0003-0064-4861}
\affiliation{%
  \institution{University of Pennsylvania}
  \city{Philadelphia}
  \country{United States}
}

\author{Hong Shui}
\email{shui0003@umn.edu}
\affiliation{%
  \institution{University of Minnesota}
   \city{Minneapolis}
  \country{United States}
}

\author{Bodong Chen}
\email{cbd@upenn.edu}
\affiliation{%
  \institution{University of Pennsylvania}
   \city{Philadelphia}
  \country{United States}
}

\renewcommand{\shortauthors}{Zhu, Shui, and Chen}

\begin{abstract}
  The ability to synthesize information has emerged as a critical skill for success across various fields. However, within the field of education, there is a lack of systematic understanding and well-defined design infrastructures that address the mechanisms and processes of knowledge synthesis in collaborative learning settings. In this poster, we introduce a design innovation – {\itshape The Synthesis Lab}, which aims to support students in synthesizing ideas from their online discussions in higher education classrooms. The tool offers structured work-spaces for students to decompose the synthesis process into intermediate synthesis products and features two key iterative processes of knowledge synthesis in collaborative settings: {\itshape categorizing peers’ ideas into conceptual building blocks} and {\itshape developing a synthesis of the discussions}. Future implementation and evaluation of the design will make significant contributions to both research and practice. 
\end{abstract}

\ccsdesc[500]{Human-centered computing~Collaborative and social computing systems and tools}
\ccsdesc[500]{Applied computing~Collaborative learning}

\keywords{CSCL, Knowledge Synthesis}


\maketitle

\section{Introduction}
In our ever-evolving world, where information flows incessantly amidst unprecedented technological advancements and the growth of artificial intelligence, the ability to synthesize information has emerged as a critical skill for success across various fields. Just as a scientist combines different reactants in a chemical experiment to create new substances, or a composer weaves melodies into a harmonious symphony, knowledge synthesis can be seen as both an art and a science. It involves skillfully and strategically weaving together diverse strands of information to foster conceptual innovation, generate novel knowledge, and design creative solutions \cite{deschryverHigherOrderThinking2014, nonakaKnowledgeCreatingCompany1998, qianOpeningBlackBox2020, scardamaliaKnowledgeBuildingKnowledge2014a}. 

Knowledge synthesis is one important form of cognition in human learning and collaboration. In contrast to other cognitive processes such as interpreting and evaluating new information, synthesis-making requires efforts to rise above current levels of explanation which results in understanding phenomena on a higher plane and the creation of new concepts \cite{vanaalstDistinguishingKnowledgesharingKnowledgeconstruction2009a, scardamaliaKnowledgeBuildingKnowledge2014a}. Within organizations or learning communities, the synthesis-making process becomes even more intricate when individuals engage in dynamic interactions, encountering a broad range of perspectives and resources, which all make the synthesis process challenging. 

Research from various disciplines has examined processes or concepts related to knowledge synthesis in collaborative/cooperative settings from multiple perspectives. In CSCW, scholars have emphasized key components of scholarly knowledge synthesis, such as capturing context information and information reuse \cite{ackermanOrganizationalMemoryObjects2004a, morabitoManagingContextScholarly2021a}. Similarly, in creativity research, researchers have used constructs that are closely related. Javadi and Fu \cite{javadiIdeaVisibilityInformation2011} investigated “idea integration” in electronic brainstorming as a process for “adoption, exploitation, combination or synthesis” of multiple ideas. In information sciences, Robert et al. \cite{robertjr.SocialCapitalKnowledge2008} refers to “knowledge integration” as “the synthesis of individual team members’ information and expertise through 'social interactions'.” These studies have shed light on the importance of effective knowledge synthesis in enhancing collaboration and improving outcomes in diverse domains. 

Moreover, educational research has also touched upon concepts related to knowledge synthesis. DeSchryver \cite{deschryverHigherOrderThinking2014} developed a framework for web-mediated knowledge synthesis which includes six strategies for individuals such as divergent keyword search, synthesis for meaning, in-the-moment insights, repurposing (e.g., engaging learners to evolve the original ideas with their own added value), reinforcement (e.g., justifying new ideas by revisiting the sources or further discussion with peers), and note-taking. Recent work framed synthesis as a “trans-disciplinary skill” that “encapsulate the ways in which creative people think.” \cite{TransformTranscendSynthesis} In CSCW’s sister field – Computer-Supported Collaborative Learning (CSCL), knowledge synthesis plays a crucial role in collaborative learning by helping students distill, connect, organize, and analyze the information to deepen their thinking. For example, the knowledge building model emphasizes the notion of “rise above” in knowledge building discourse to synthesize and build on previous ideas that leads to the development of novel knowledge \cite{scardamaliaKnowledgeBuildingKnowledge2014a}. However, there is a lack of systematic understanding regarding the mechanisms and processes of knowledge synthesis in CSCL. Important questions remain unanswered, necessitating both theoretical and empirical investigation in the field. For instance, {\itshape how do students synthesize ideas generated in collaborative discourse? How can the knowledge synthesis process support learning and collaboration?} And {\itshape how can the synthesis be used to orchestrate various learning events?} Such understanding is essential for informing the design of learning systems, technologies, and pedagogies to support effective knowledge synthesis. 

To address this gap, we initiated a Design-Based Research \cite{barabDesignBasedResearchPutting2004} project, connecting theories and designs in CSCW and CSCL, to understand and support knowledge synthesis through a series of ongoing design innovations. Drawing on the socio-cognitive stances, this project aims to 1) support the knowledge synthesis process in CSCL through a series of design innovations, and 2) investigate the mechanism of knowledge synthesis in collaborative learning settings through empirical research. In this poster, we aim to showcase an early effort of this project, the design of a web application for supporting knowledge synthesis in college students’ online discussion activities – {\itshape The Synthesis Lab}. This application helps deconstruct the complex synthesis-making process into smaller building blocks and guides students through the key steps, including distilling, connecting, analyzing, rising above, and aggregating ideas generated from the discussions. These steps guide students to discover the interrelationship between peers’ posts other than the simply reply relationships, which leads to further rising above previous ideas and constructing coherent knowledge out of fragmentary information.

\section{THE SYNTHESIS LAB}

Informed by interdisciplinary literature, knowledge synthesis in the designed technology space is operationalized as a dynamic process encompassing the analysis and integration of ideas fostered through interactions with peers in digital environments. Situated in a collaborative setting, the overarching goal is to generate novel knowledge out of the conversation, while facilitating the orchestration of various learning activities and application scenarios. Serving as a tool for thinking, it nurtures higher order competences, such as creativity and collaboration, fostering a fertile ground for profound thinking and intellectual growth.

\begin{figure*}
  \centering
  \includegraphics[width=\textwidth]{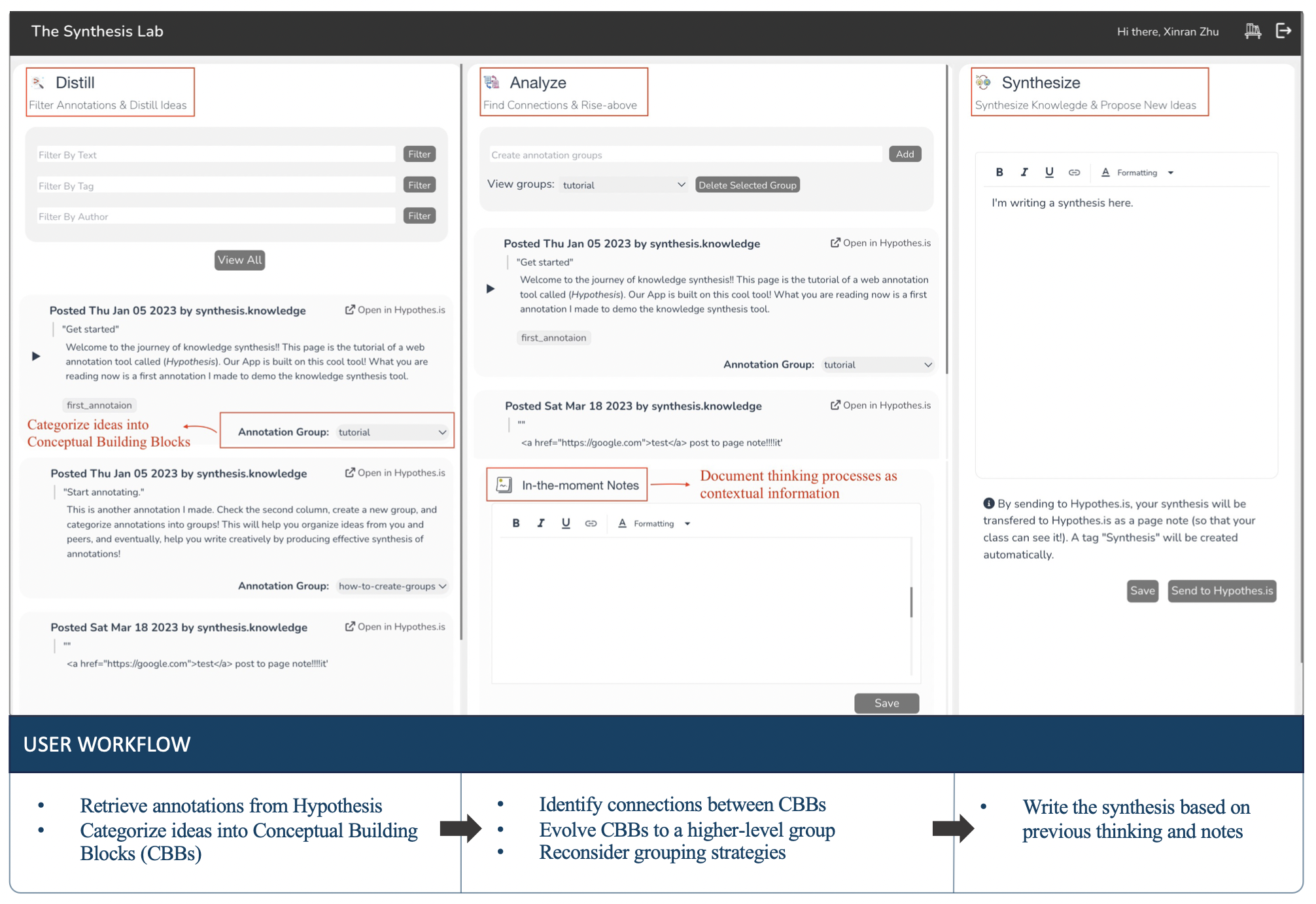}
  \caption{The Interface and Example User Workflow of The Synthesis Lab. (\url{https://h-synthesizer.web.app/}).} 
  \Description{Interface and workflow.}
  \label{fig:interface}
\end{figure*}

The Synthesis Lab (see Fig.~\ref{fig:interface} for the interface and example user workflow) retrieves students’ online discussion data on a web annotation platform – {\itshape Hypothesis} (https://web.hypothes.is/), via its APIs. Hypothesis is a web annotation technology that allows users to collaboratively read, annotate, highlight, and tag on a shared document or web page. It has been widely used to support social reading in classrooms as a form of online discussion across universities \cite{zhuReadingConnectingUsing2020}. For example, as part of their weekly routine, students engage with course readings by annotating the texts and responding to peers’ annotations prior to in-person class meetings. 

Drawing inspiration from previous designs (e.g., \cite{chanSustainableAuthorshipModels2021,hahnKnowledgeAcceleratorBig2016a}) and incorporating insights from interdisciplinary literature (e.g., \cite{ackermanSharingKnowledgeExpertise2013, bloom, deschryverHigherOrderThinking2014, morabitoManagingContextScholarly2021a, robertjr.SocialCapitalKnowledge2008}), The Synthesis Lab offers a structural framework to guide students’ synthesis process. The workflow within the tool revolves around two primary goals: {\itshape categorizing peers’ ideas into conceptual building blocks (CBBs)} \cite{morabitoManagingContextScholarly2021a} and {\itshape developing a synthesis of the discussions}. These goals are achieved through interaction across three vertically organized workspaces: {\itshape Distill}, {\itshape Analyze}, and {\itshape Synthesize}. This organization provides structured workspaces for students to decompose the tasks into intermediate synthesis products: in-source annotations, per-source summaries, and cross-source syntheses \cite{qianOpeningBlackBox2020}. The design encourages students to fluidly navigate between these workspaces, allowing them to revisit annotations and thoughts iteratively, recognizing that the synthesis-making process is non-linear in nature. 

\subsection{Categorizing Peers’ Ideas into CBBs}

Once students have selected the reading for analysis, they initiate the synthesis process by browsing through the class annotations. In the Distill column, students are able to filter annotations by key words, authors and tags. Meanwhile, 
students start to analyze annotations by creating Annotation Groups in the Analyze column, where they categorize annotations into different categories following on various strategies. For instance, some students may opt to group annotations by “applications” or “methodology”, while others may group them based on semantic meanings. This step allows students to organize ideas into CBBs, which become the metadata and contextual information for future synthesis work \cite{morabitoManagingContextScholarly2021a}. Additionally, students jot down their thoughts in the "In-the-moment Notes" box to document the contextual information surrounding their decisions. This step encourages active analysis of peers’ ideas and the meaningful integration of concepts.

\subsection{Developing a Synthesis of the Discussions} 
Following their analysis of individual annotations, this step prompts students to shift their attention to the Annotation Groups in order to identify connections or reconsider their grouping strategies. For example, they can merge two groups as a new group (combining CBBs to a higher level CBB) or transfer annotations from one group to another. This process encourages students to repurpose and reinforce their learning by ruminating over the categories and revisiting the annotations/notes \cite{deschryverHigherOrderThinking2014}. Ultimately, students start the synthesis writing phase in the Synthesize column, drawing upon all their existing notes and activities to compose a comprehensive synthesis. 

\section{CURRENT IMPLEMENTATION AND FUTURE DIRECTIONS}
Using a co-design approach, we are collaborating with instructors to develop class activities that were supported by The Synthesis Lab. To investigate the design enactment, we conducted a pilot study during the Spring 2023 semester in a graduate-level classroom at a large private university. Throughout the study, students actively engaged in social annotation activities on a weekly basis. Within each session, 2 to 3 students assumed the role of discussion leaders, facilitating in-class discussions. To effectively fulfill their role, the discussion leaders were required to synthesize the class annotations in preparation for the meetings. For this study, discussion leaders who volunteered to participate utilized The Synthesis Lab to support their synthesis process. 

We collected various learning artifacts from the participants, including their annotations and synthesis writings. Additionally, we conducted follow-up interviews with three participants to gain further insights. Upon preliminary analysis of the collected data, we noticed that the synthesis strategies employed by students varied. For instance, one student adopted a deductive approach by initially creating Annotation Groups based on the abstract, and then assigning relevant annotations to these pre-defined groups. Conversely, another student employed an inductive approach, generating CBBs while reading annotations within the Distill space in an iterative manner. Furthermore, the analysis highlights the potential of this design in fostering a more profound understanding of the value of knowledge synthesis among students. It also demonstrates the capacity of the design to enhance students' synthesis skills and promote collaborative learning.

The first design iteration is currently in progress and is expected to be completed by Fall 2023. Our focus is on enhancing the interactivity between different workspaces, including the implementation of backlinks to maintain the connection between annotations and In-the-Moment Notes. Additionally, we are actively exploring how artificial intelligence (AI) can augment the synthesis process to further enhance the students' experience. 

The expected contribution of this work will be three-fold. First, the proposed technology innovation has great potential for broader applications. Further, the investigation of the design's implementation aims to develop a framework of knowledge synthesis in collaborative learning that will make significant contributions to both research and practice. Finally, leveraging the synergies between CSCW and CSCL allows for a deeper understanding of the interplay between technology, social dynamics, and learning constructs within the knowledge synthesis process. This understanding, in turn, allows for the creation of meaningful design infrastructures that can contribute to both fields advancing our understanding of learning and collaboration processes, optimizing technology-supported interactions, and fostering creative knowledge creation.

\bibliographystyle{ACM-Reference-Format}
\bibliography{sample-base}

\end{document}